\documentclass[aps,pra, twocolumn]{revtex4}

\usepackage{graphicx}
\usepackage{amssymb}
\usepackage{amsmath}
\usepackage{colordvi}
\usepackage{color}

\newcommand{\PT}{{\cal PT}}

\newcommand{\cL}{{\cal L}}
\newcommand{\cZ}{{\cal Z}}

\newcommand{\ta}{\tilde{a}}

\newcommand {\sgn}{\textrm{sgn\,}}

\begin{document}

\title{Stability restoration by asymmetric nonlinear states in non-Hermitian double-well 	potentials}

\author{Dmitry A. Zezyulin}

\affiliation{School of Physics and Engineering, ITMO University, St. Petersburg 197101, Russia\\
Abrikosov Center for Theoretical Physics, MIPT, Dolgoprudny, Moscow Region 141701, Russia
}

\date{\today}

\begin{abstract}

We introduce a class of one-dimensional complex optical potentials that feature a nonlinearity-induced stability restoration, i.e., the existence of stable nonlinear modes propagating in a waveguide whose  linear eigenmodes are unstable. The optical potential is an even function of the transverse coordinate, i.e., the system is parity symmetric but not parity-time  symmetric. The stability restoration occurs for asymmetric  stationary nonlinear modes that do not respect the parity symmetry. Stable nonlinear states exist either for focusing and defocusing nonlinearities. On the qualitative level   the stability restoration cab be analyzed  using  a simple bimodal system. Its solutions enable  systematic construction of stable stationary modes and more complex  patterns with intensity periodically oscillating along the propagation distance.

\end{abstract}

\maketitle

\section{Introduction}

Ongoing progress  in  experimental realization  of  non-Hermitian optical systems \cite{Konotop2016,Such2016,Feng2017,El2018,Ozdemir,Gupta2019,Nasari2023} intensifies the discussions on the joint impact that non-Hermiticity and nonlinearity make  on the existence and stability of guided  waves. Either nonlinearity and non-Hermiticity make the propagation of light more intricate  and  are therefore expected to make the stability more fragile. In the meantime, a  few  remarkable  situations are known where the nonlinearity  makes the  propagation of non-Hermitian waves more  stable as compared to   effectively linear states.  A prominent development in this direction has been made by the proposal of the concept of nonlinearly  induced parity-time ($\PT$) symmetry transition \cite{Lumer}  (see also \cite{Zhang2021}). For a class of periodic $\PT$-symmetric potentials, authors  of  Ref.~\cite{Lumer}   found that if the optical power is high enough,  the Bloch spectrum of nonlinear waves can become   purely real, even though the  spectrum of  propagation constants of linear waves is partially complex. A similar    behavior was encountered for a class of bimodal $\PT$-symmetric  systems \cite{BG,Barash2015,Barash2023}, where the total power   remained bounded  (or even constant) along the propagation distance regardless of the value of the gain--loss coefficient. In a   weaker sense, the stabilization by nonlinearity  can be understood as the existence of dynamically stable nonlinear modes in a system whose   linear waves  (i.e., waves of infinitesimally small amplitude) are unstable. Using the language of spectral stability, in these systems the nonlinear potential suppresses    eigenmodes that were unstable in the underlying linear limit.  Situations of this type have been encountered multiple times in previous studies, either in finite-mode setups \cite{Zezyulin2012} and in spatially continuous waveguides  \cite{Yan2015,YanWenKonotop2015,Chen2016,Chen2017,Saha2018}. Still,   systematic studies of  the nonlinearity-induced stabilization  are  scarce and mainly limited to  $\PT$-symmetric systems.

The main goal of the present paper is to introduce a nonlinearity-induced stability restoration in a continuous and non-$\PT$-symmetric   system. We consider light propagating in  a  parity-symmetric (but not  parity-time  symmetric) complex potential whose real part has a double-well structure (or a double-hump structure if considered as an optical potential created by transversely varying     refractive index). We show that in  the linear regime the system cannot have stable localized guided modes, and instability of linear   modes  is    {protected}  by the parity symmetry of the potential. The degree of   instability of linear waves is mediated by the distance between the potential wells. Most importantly, we find that   nonlinearity     suppresses the   unstable eigenvalues and enables  stable propagation of stationary states and  periodically oscillating    patterns. In a very  unconventional  manner, the nonlinearity-induced stabilization   occurs for {asymmetric} stationary modes  which do not respect the parity symmetry inherent to the system. Another important  feature is that   stabilization occurs for either sign (focusing and defocusing) of the optical nonlinearity. 

The rest of the paper is organized as follows. In Sec.~\ref{sec:model} we introduce our model and discuss its instability in the linear regime. Section~\ref{sec:main} contains the main results on the existence of stable nonlinear patterns. In  Sec.~\ref{sec:concl} we summarize and discuss  the outcomes of our study.  

\section{The model and its linear properties}
 \label{sec:model}
We model the light propagation using the following dimensionless equation  for complex-valued \textcolor{black}{envelope} of the electric field $\Psi(x,z)$:
\begin{eqnarray}
\label{eq:nls}
i\Psi_z +  \Psi_{xx}  + [w^2(x)  + i w_x(x)]\Psi  + \sigma |\Psi|^{2}\Psi =0.
\end{eqnarray} 
In this nonlinear Schr\"odinger-like equation $z$ and $x$ stay for longitudinal and transverse coordinates, and $\sigma$ is a coefficient of   cubic Kerr nonlinearity. We consider   both  focusing ($\sigma>0$) and defocusing ($\sigma<0$) cases. Equation (\ref{eq:nls}) incorporates an  additional optical potential that describes the modulation of   complex-valued refractive index in the transverse direction. The corresponding landscape is given by the term  $w^2(x) + iw_x(x)$, where  $w(x)$ is a smooth real-valued function, $w_x(x)$ is its derivative, and $i$ is the imaginary unit. Potentials of these form are sometimes referred to as Wadati potentials, after the author of Ref.~\cite{Wadati}  where their  relevance was emphasized in the context of $\PT$ symmetry. 

To explain  such a choice of the optical landscape, we note that potentials of this form have been already used to construct continuous families of nonlinear modes and solitons in   $\PT$-symmetric and asymmetric non-Hermitian waveguides \cite{Tsoy,KonZez14,quartets,Yang20,saturable,Zezyulin2023}. In the $\PT$-symmetric case [which corresponds to even functions $w(x) = w(-x)$], Wadati potentials enable pitchfork $\PT$-symmetry breaking bifurcations \cite{Yang2014} that are unknown to occur in $\PT$-symmetric potentials of   other shapes. Another important feature of Wadati potentials is that their linear eigenvalues are either real or exist in complex-conjugate pairs \cite{NixonYang16PRA}. Therefore, similar to $\PT$-symmetric systems,   Wadati potentials can undergo the phase transition  from all-real to partially complex spectra \cite{Yang17,ZK20}. A possibility of  optical realization of Wadati potentials has been suggested for light propagating in a gas of  coherent multilevel  atoms driven by an external laser field \cite{Hang2017}.   

Regarding the stability restoration discussed in the present paper, it becomes possible due to special properties of  Wadati potentials explained below.  In contrast to most of   previous studies dealing with   Wadati potentials,    we consider $w(x)$ to be an antisymmetric (i.e., odd) function: 
\begin{equation}
w(x) = -w(-x).
\end{equation}
We also assume that $w(x)$ and its derivative $w_x(x)$  decay rapidly    as $x\to+\infty$ and $x\to-\infty$. For such a choice of function $w(x)$, the resulting optical potential $w^2(x)  + i w_x(x)$ is parity symmetric  but \textit{not} parity-time symmetric (nevertheless, as we will discuss below, its properties can be, in a certain sense, approximated by a $\PT$-symmetric model).

Our first result applies to the linear case [i.e., $\sigma=0$ in Eq.~(\ref{eq:nls})]. We argue that   a linear waveguide with the potential introduced above   can not guide localized eigenmodes with real propagation constants. Indeed, for stationary modes $\Psi = e^{i\beta z} \psi(x)$,  Eq.~(\ref{eq:nls}) with $\sigma=0$ becomes a     linear  eigenvalue problem:
\begin{equation}
\label{eq:linear}
\beta \psi = \cL \psi, \qquad \cL  = \partial_x^2 + w^2(x)  + i w_x(x),
\end{equation}
where $\beta$ is the propagation constant. Operator $\cL$ in   Eq.~(\ref{eq:linear})  is   a non-self-adjoint one-dimensional Schr\"odinger operator (precisely speaking,  $-\cL$ is a  Schr\"odinger operator). Since potential  $w^2(x)  + i w_x(x)$ rapidly decays to zero, operator $\cL$ has a continuous spectrum which occupies the semiaxis $\beta \in (-\infty, 0]$. Operator $\cL$ can also have discrete eigenvalues associated with localized eigenfunctions $\psi(x)$. These eigenvalues (if any) are either   positive real numbers or complex numbers.   Due to the parity symmetry, any localized eigenfunction $\psi(x)$ of  operator $\cL$ in (\ref{eq:linear})  is either   even  or odd function of $x$. On the other hand, from the properties of Wadati potentials it is known \cite{NixonYang16PRA} that   if $\psi(x)$ is an eigenfunction  corresponding to an eigenvalue $\beta$, then $(\eta \psi(x))^*$, where $\eta = \partial_x + iw(x)$, is an eigenfunction corresponding to eigenvalue $\beta^*$ (the asterisk means complex conjugation). \textcolor{black}{If $\beta$ is real, then two eigenfunctions $\psi$ and $(\eta \psi(x))^*$ correspond  to the same eigenvalue. However,  it is well-known that in one dimension the Schr\"odinger operator cannot have multiple eigenvalues with localized eigenfunctions \cite{LL}. Hence  $\psi$ and $(\eta \psi(x))^*$ must be linearly dependent, which is   impossible, because  functions $\psi(x)$ and $\eta \psi(x)$ have opposite parities. Therefore the propagation constant    $\beta$ cannot be real.} 

\textcolor{black}{The same    result can also be deduced from earlier literature. To this end, let us recall the well-known connection between the eigenvalue equation (\ref{eq:linear}) and the Zakharov-Shabat (ZS) spectral problem \cite{ZS}  that plays the central role in   the inverse scattering theory for the modified Korteweg-de Vries   equation \cite{Lamb}. Indeed, let us consider the ZS problem in the form
\begin{equation}
p_x = -i\zeta p + w(x) q, \quad q_x =  i\zeta q - w(x) p,
\end{equation}
for  eigenvector  $(p(x), q(x))$ and eigenvalue   $\zeta$.    Then function $\psi =  q - ip$ satisfies Eq.~(\ref{eq:linear}) with $\beta = -\zeta^2$ \cite{Lamb}; see also \cite{Tsoy,KonZez14,Konotop2016}. It is known that if $w(x)$ decays to zero rapidly enough, then the continuous spectrum of the ZS problem fills the whole real axis $\zeta \in \mathbb{R}$. At the same time, the ZS problem can also have discrete eigenvalues $\zeta$ that are always situated symmetrically with respect to the imaginary axis, i.e., if $\zeta$ is an eigenvalue, then so is $-\zeta^*$. Guided modes with localized   envelopes $\psi(x)$ and  real propagation constants $\beta$ correspond to purely imaginary eigenvalues $\zeta$ of the ZS problem. However, according to   Theorem~3.4 in Ref.~\cite{Klaus2003},  if $w(x)$ is an odd function, then the ZS problem has no purely imaginary eigenvalues.}

\textcolor{black}{As a sidenote,   Wadati potentials  $ w^2(x)  + i w_x(x)$ are strongly reminiscent of the famous Miura transformation  \cite{Miura} that relates the  Korteweg-de-Vries   equation and the  modified Korteweg-de Vries   equation.}

We have established that  for   antisymmetric   function $w(x)$  the discrete spectrum of propagation constants  is either empty or consists of one or several   complex-conjugate  pairs with  eigenfunctions  having opposite parities and being  intertwined  by    operator $\eta$. In each complex-conjugate pair, one of the  modes   indefinitely grows along the propagation distance, which renders the linear waveguide unstable.
Changing the sign of function $w(x)$  complex-conjugates  the spectrum, i.e., attenuated and amplified   modes swap.

\begin{figure}
	\begin{center}		
		\includegraphics[width=0.999\columnwidth]{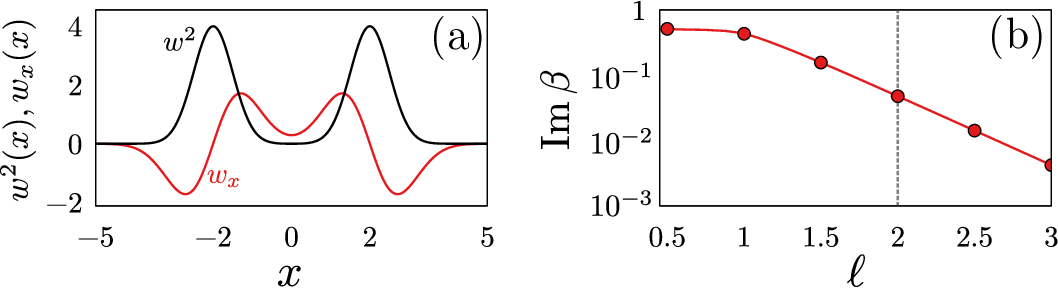}	
		\caption{(a) Real [$w^2(x)$] and imaginary [$w_x(x)$] parts of an optical potential generated by function (\ref{eq:exp}) with $W_0=2$ and $\ell = 2$. (b) Imaginary part of the unstable propagation constant  for $W_0=2$ and increasing well separation $\ell$.  		\label{fig:linear} }
	\end{center}
\end{figure}

In the rest of this paper we consider the situation  where the real part of the optical potential has a double-lobe structure. A representative example of suitable   odd function $w(x)$ is
\begin{equation}
\label{eq:exp}
w(x) = W_0e^{-(x-\ell)^2} - W_0e^{-(x+\ell)^2},
\end{equation}
where $W_0$ is the amplitude and $\ell$ is the halfdistance between the lobes, see  plot in  Fig.~\ref{fig:linear}(a). As the halfdistance $\ell$ increases, the complex-conjugate propagation constants become closer to the real axis, i.e., the instability weakens [see Fig.~\ref{fig:linear}(b)]. For sufficiently large $\ell$ the decay of  imaginary parts is nearly exponential. Our main result demonstrates that   even in the regime where the instability increment of the  linear propagation constant    is not yet very small, the   nonlinearity       enables propagation of stable modes.

\section{Main results}
 \label{sec:main}
\subsection{Two-mode system}

\textcolor{black}{We consider the situation where the linear operator $\cL$ in Eq.~(\ref{eq:linear})  has exactly two discrete propagation constants which are denoted as $\beta_\pm$.} According to the above analysis, these values must be complex-conjugate: $\beta_+ = \beta_-^*$. We denote the corresponding   eigenfunctions   as  $\psi_\pm(x)$ and normalize them such that   $\int_{-\infty}^\infty |\psi_\pm(x)|^2dx = 1$. As explained above, these functions have opposite parities and therefore   orthogonal: $\int_{-\infty}^\infty \psi_+^*(x)\psi_-(x)dx = 0$. This   distinguishes our analysis from the $\PT$-symmetric case, where the eigenfunctions associated with different eigenvalues are  generically  not orthogonal in the usual sense.

We further assume that phases of   eigenfunctions $\psi_\pm(x)$ are chosen such that the following linear combinations  corresponds to the superpositions localized in the right (R) and left (L) wells:
\begin{equation}
\varphi_{R,L}(x) =  [\psi_+(x) \pm \psi_-(x)]/\sqrt{2}.
\end{equation}
These functions are  normalized and  orthogonal  too:  $\int_{-\infty}^\infty \varphi_R^*(x)\varphi_L(x)dx = 0$, $\int_{-\infty}^\infty |\varphi_{R,L}(x)|^2dx= 1$.
We introduce a pair of real constants
\begin{equation}
\label{eq:varsigma}
b = \frac{\beta_+ + \beta_-}{2}, \quad \gamma = \frac{\beta_- - \beta_+}{2i}.
\end{equation}
Without loss of generality $\gamma>0$.  We look for  solutions  of Eq.~(\ref{eq:nls}) in the form of a  two-mode substitution  
\begin{equation}
\label{eq:Psi}
\Psi(x,z) = e^{i b z}  [a_R(z) \varphi_{R}(x) + a_L(z) \varphi_{L}(x) ].
\end{equation} 
If   the separation between the well (resp., humps) of the   potential is large enough,i.e., $2\ell \gg 1$ in Eq.~(\ref{eq:exp}), then  the integrals containing the  crossproducts between  $\varphi_{R}$ and $\varphi_{L}$ can be assumed  negligible, and one can approximate the evolution of the envelope $\Psi(x,z)$ by   the following   coupled-mode system
\begin{equation}
\label{eq:aRaL}
 \begin{array}{rcl}
i\dot{a}_R &=& i\gamma a_L - \chi |a_R|^{2} a_R,\\[1mm]
i\dot{a}_L &=& i\gamma a_R - \chi |a_L|^{2} a_L,
\end{array}
\end{equation}
where overdots mean   derivatives with respect to $z$, and  $\chi  =  \sigma \int_{-\infty}^\infty |\varphi_R|^{4}dx =  \sigma \int_{-\infty}^\infty |\varphi_L|^{4}dx$.

Even though  system (\ref{eq:aRaL})  is obtained using the standard two-mode approach, it  differs dramatically from the previously considered systems describing bosonic Josephson osillations in real double-well potentials (see e.g. \cite{Ana1996,Sacc2009}) due to the  imaginary unit in the coupling terms. This change renders the  dynamics   non-Hermitian. In particular,   the imbalance between the  right  and left  amplitudes  is conserved:
\begin{equation}
\label{eq:notunnel}
\partial_z (|a_R|^2 - |a_L|^2) = 0.
\end{equation}
This is in a stark contrast to  the periodic tunneling between the left and right subsystems   typical of  usual   double-well (or double-lobe) potentials. 

\textcolor{black}{In the linear case (that is for $\chi=0$), the general   solution of  system (\ref{eq:aRaL}) is
\begin{equation}
\left(
\begin{array}{c}
a_R\\a_L
\end{array}
\right) = \alpha_1 e^{\gamma z} \left(
\begin{array}{c}
1\\1
\end{array}
\right) + \alpha_2 e^{-\gamma z}   \left(
\begin{array}{c}
1\\-1
\end{array}
\right),
\end{equation}
where $\alpha_{1,2}$ are constants.  The solution corresponding to the growing exponent  $ e^{\gamma z}$  blows up to infinity as $z\to\infty$. Therefore   the zero equilibrium  $(a_R, a_L) = (0, 0)$ is an unstable fixed point of the nonlinear system (\ref{eq:aRaL}).}

\textcolor{black}{ 
After a simple transformation, the coupled-mode system (\ref{eq:aRaL}) can be brought to the form of  a $\PT$-symmetric nonlinearly coupled   dimer.   Representing  $a_R =  u+v$,   $a_L =  u-v$, where $u$ and $v$ are   new functions, from system   (\ref{eq:aRaL})  we obtain
\begin{equation}
\label{eq:cubic}
\begin{array}{lcr}
i\dot{u} &=& \phantom{+}i\gamma u  -\chi[(|u|^2+2|v|^2)u + v^2u^*],\\[2mm]
i\dot{v} &=& -i\gamma v - \chi[(|v|^2+2|u|^2)v + u^2v^*].
\end{array}
\end{equation}
A very similar (yet not fully identical) dynamical system has been earlier considered in \cite{BG}  as an  exactly solvable $\PT$-symmetric dimer. For that system    the authors of Ref.~\cite{BG}   found that the nonlinearity ``softens''  the $\PT$  phase transition, i.e.,  stable nonlinear states can be found for arbitrarily large  non-Hermiticity coefficient $\gamma$, provided that the  solution amplitude  is large enough.  In contrast to our system in Eqs.~(\ref{eq:cubic}), the $\PT$-symmetric dimer  considered in Ref.~\cite{BG} contained  additional   linear coupling, i.e., terms $\kappa v$ and $\kappa u$ in the first and second equations, respectively, where $\kappa>0$ was the coupling coefficient. In our system, that linear coupling is absent.  This difference leads to  a moderate modification of our two-mode analysis as compared to that developed in Ref.~\cite{BG}.}

Dynamics of the bimodal system (\ref{eq:aRaL}) can be analyzed   in terms of the  Stokes parameters  defined as
\begin{eqnarray}
\{X,Y,Z\} = \left( \begin{array}{cc}
a_R^*, & a_L^*
\end{array}\right)
\sigma_{\{x,y,z\}} \left( \begin{array}{cc}
a_R, & a_L
\end{array}\right)
^T,
\end{eqnarray} 
where $\sigma_{x,y,z}$ are the Pauli matrices, and superscript $T$ means transposition.  In those variables system  (\ref{eq:aRaL}) transforms into  
\begin{equation}
\label{eq:Stokes}
\begin{array}{rcl}
\dot{X} &=& 2\gamma A + \chi Y Z,\\[1mm]
\dot{Y} &=& -\chi XZ,\\[1mm]
\dot{Z} &=& 0.
\end{array}
\end{equation}
These equations additionally involve   the length of the Stokes vector $A$ which   characterizes the total energy stored in  both modes:
\begin{equation}
A =  (X^2 + Y^2 + Z^2)^{1/2} =   |a_R|^2 + |a_L|^2.
\end{equation}
Its dynamics obeys the following equation:
\begin{equation}
\label{eq:dotA}
\dot{A} = 2\gamma X.
\end{equation}

Equations (\ref{eq:Stokes}) and  (\ref{eq:dotA})  can be combined into a  simple linear equation:
\begin{equation}
\ddot{X} + \nu^2 X=0,
\end{equation}
where we have introduced 
\begin{equation}
\label{eq:nu2}
\nu^2 =  (\chi Z)^2 - 4\gamma^2.
\end{equation}
Bounded orbits are possible only  if the condition $\nu^2>0$ is satisfied, that is for $|\chi Z| > 2\gamma$.  Since   the Stokes parameter $Z=|a_R|^2 - |a_L|^2$ corresponds  to the imbalance between    the amplitudes of right and left modes, we conclude that, in terms of the starting equation (\ref{eq:nls}),  any stable solution (if exists at all)  \textit{must} be asymmetric with respect to the parity reversal: $|\Psi(x,z)| \ne |\Psi(-x, z)|$.

 
System  (\ref{eq:Stokes}), (\ref{eq:dotA}) is fully solvable \cite{BG}. For $\nu^2>0$  the general solution reads  
\begin{eqnarray}
\label{eq:Xz}
X(z) &=& \rho_0\cos\phi,\\[1mm]
\label{eq:Yz}
Y(z) &=& \frac{-1}{\nu}\left[2\gamma    \sqrt{Z^2 + \rho_0^2}\, \sgn (\chi Z) +  \chi Z \rho_0 \sin \phi\right],\\[1mm]
Z &=& \mathrm{const},
\end{eqnarray}
where $\phi = \nu(z-z_0)$, and $z_0$ and $\rho_0$ are arbitrary constants of integration.  The value  of the conserved quantity $Z$ is   arbitrary too,  provided that   inequality $\nu^2>0$ holds. For  the length of the Stokes vector $A$   we compute
\begin{equation}
A(z) =\frac{1}{\nu}\left[ {\sqrt{Z^2 + \rho_0^2}} |\chi Z| + 2 \gamma \rho_0 \sin \phi\right].
\end{equation}

If $\nu^2<0$, all nontrivial  solutions of system  (\ref{eq:Stokes}), (\ref{eq:dotA})  grow indefinitely as $z\to+\infty$. We do not consider this case.

\subsection{Fixed points of the two-mode system}

\begin{figure}
	\begin{center}		
		\includegraphics[width=0.999\columnwidth]{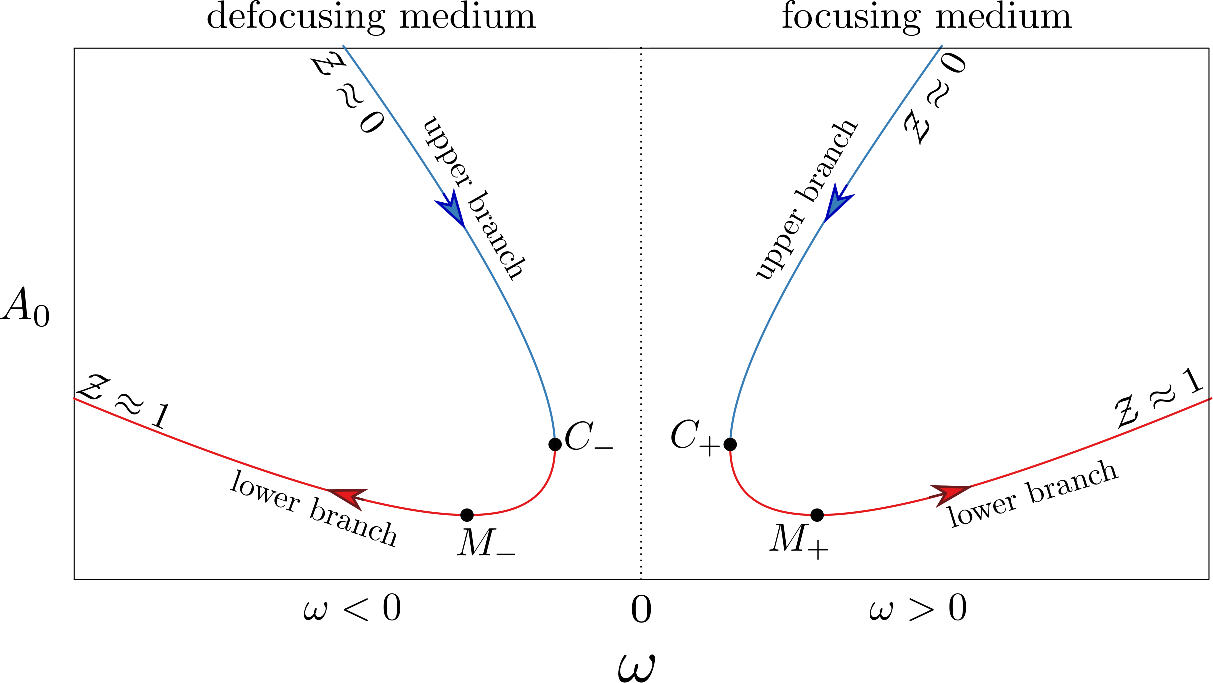}	
		\caption{\textcolor{black}{Schematic diagram of fixed points of the two-mode system (\ref{eq:aRaL})  displayed as dependencies   $A_0(\omega)$  in focusing and defocusing media.  In terms of the full system, quantities $A_0$ and $\omega$ transform in  squared norm  and   propagation constant of stationary nonlinear modes,  respectively; compare this  figure with  Fig.~\ref{fig:families} which shows   analogous dependencies produced from full envelope equation~(\ref{eq:nls}).  Arrows show   directions which correspond  to the increase of the normalized asymmetry parameter   $\cZ$ from its asymptotic minimal value  $\cZ=0$ up to  the maximal asymptotic value $\cZ = 1$. Labels  $C_\pm$ mark the points where the upper and  the lower subbranches meet pairwise.    Labels  $M_\pm$ mark the points where the minimal values of $A_0$ is achieved. } 
			\label{fig:schematics} }
	\end{center}
\end{figure}

The conserved quantity $Z$ can be  used to parameterize   found solutions in the phase  space $(X,Y,Z)$. Any  motion is   restricted to the plane $Z=\mathrm{const}$, and    every solution $(X(z), Y(z))$ given by   Eqs.~(\ref{eq:Xz})--(\ref{eq:Yz})  forms an ellipse in the corresponding $Z$-plane.  Moreover,  each   plane  $Z=\mathrm{const}$ is filled by nested ellipses that correspond  to different values of the integration constant $\rho_0$ in  Eqs.~(\ref{eq:Xz})--(\ref{eq:Yz}).  Each plane  $Z=\mathrm{const}$ contains   exactly one fixed point  which can be found by setting    $\rho_0=0$:
\begin{eqnarray}
\label{eq:X0}
X_0 = 0, \quad Y_0  = -2\nu^{-1} \gamma  Z \, \sgn \chi,   \quad A_0  =  \nu^{-1}  |\chi| Z^2.
\end{eqnarray}
\textcolor{black}{Hereafter we use subscript $0$ to distinguish values of  Stokes parameters $X$, $Y$ and $A$ that pertain to fixed points.}
This fixed point lies inside     all nested ellipses hosted in the corresponding $Z$-plane and is therefore   stable. Each fixed point corresponds to a solution of the form $a_R(z) = U_R e^{i\omega z}$, $a_L(z) = U_L e^{i\omega z - i\alpha}$, 
where $U_{R,L}$, $\alpha$ and $\omega$   are real   constants. Value of $\omega$    can be retrieved from the two-mode system  (\ref{eq:aRaL}):
\begin{equation}
\label{eq:omega}
\omega = \chi\frac{A_0^2+Z^2}{2A_0} = \sgn \chi \frac{\chi^2 Z^2 - 2\gamma^2}{\sqrt{\chi^2 Z^2-4\gamma^2}}.
\end{equation} 
\textcolor{black}{Parameter $\omega$  is   physically relevant  because it gives the distance between the propagation constant $\beta$ of a nonlinear mode and the     point, where the linear modes are situated, i.e., from $b = (\beta_+ + \beta_-)/2$, see  Eqs.~(\ref{eq:varsigma}) and  (\ref{eq:Psi}):
\begin{equation}
\label{eq:om}
\beta = b + \omega.
\end{equation}}
It is therefore  natural to scan Eq.~(\ref{eq:omega}) for different $\omega$ and solve it with respect to  $Z$. Simple analysis shows that Eq.~(\ref{eq:omega}), considered as an equation with respect to $Z$, has real roots if and only if
\begin{equation}
\label{eq:omegac}
\sgn\omega = \sgn \chi, \qquad \mbox{and}\quad |\omega|\geq \omega_{C} := 2\sqrt{2}  \gamma,
\end{equation}
where subscript `$C$' refers to points $C_\pm$ in Fig.~\ref{fig:schematics} (see   discussion below). If conditions (\ref{eq:omegac}) hold, then  all possible solutions are given as
\begin{equation}
\label{eq:X_from_omega}
Z = \pm \frac{1}{\sqrt{2}|\chi|} \sqrt{\omega^2 + 4\gamma^2   \pm  |\omega| \sqrt{\omega^2 - \omega_{C}^2}},
\end{equation}
where  all  four combinations of  signs $+$ and $-$ are possible. If $|\omega|$ is strictly larger than the critical value $\omega_{C}$, then there are four different solutions. In fact, the analysis    can be limited only to two positive roots $Z$, because    negative $Z$    correspond to the parity reversal   $a_R \leftrightarrow a_L$.

In terms of the full model  (\ref{eq:nls}), Eq.~(\ref{eq:X_from_omega}) predicts  that   stable asymmetric nonlinear modes birth as the propagation constant $\beta$ gets far enough from the point $b$, where the unstable linear modes are situated. The critical difference between $\beta$ and $b$ is given by $\pm \omega_{C}$; this difference     must be    positive (negative) for focusing (defocusing) nonlinearity. It also readily follows from (\ref{eq:X_from_omega}) that for either sign of   nonlinearity,  two  physically distinct  nonlinear solutions are born   simultaneously at a fold-like bifurcation corresponding to $\omega = \pm \omega_{C}$.

Equations~(\ref{eq:X0}) and (\ref{eq:X_from_omega}) provide all the necessary information to   compute all Stokes parameters corresponding to  the fixed points  and trace their dependence on \textcolor{black}{the propagation constants mismatch} $\omega$. In Fig.~\ref{fig:schematics} we show a representative schematics that illustrates the behavior of the Stokes parameters  on the plane $A_0$ vs. $\omega$. \textcolor{black}{We choose these parameters to display because  they can be easily transformed to   physically relevant squared norm and propagation constant of nonlinear modes. Indeed, if $\Psi(x,z) = e^{i\beta z} \psi(x)$ is a stationary nonlinear mode, then  the two-mode substitution (\ref{eq:Psi}) implies that  $\int_{-\infty}^\infty |\psi(x)|^2dx = A_0$ and $\beta =  \omega +b$, see   Eq.~(\ref{eq:om})}. Diagram in Fig.~\ref{fig:schematics} presents   two curves: one for a  focusing and another one for a defocusing medium, and each curve   consists of two subbranches (upper branch and lower branch) which merge pairwise exactly at $\omega = \pm \omega_{C}$, where $\omega_{C}$ is the threshold value defined in (\ref{eq:omegac}).  In Fig.~\ref{fig:schematics}, we  use labels $C_\pm$ to highlight the points where solutions from upper and lower subbranches meet pairwise: 
\begin{equation}
\label{eq:merger}
C_\pm: \quad A_{0,C} = \frac{3\sqrt{2}\gamma }{|\chi|}, \qquad \omega_{C_\pm} =    \pm 2\sqrt{2} \gamma.
\end{equation}

\textcolor{black}{The diagram in Fig.~\ref{fig:schematics} also contains two other special points denoted as $M_\pm$. They   correspond to the minimal possible value of $A_0$:
\begin{equation}
\label{eq:M}
M_\pm: \quad	A_{0,M} = \frac{4\gamma}{|\chi|},  \qquad \omega_{M_\pm} = \pm 3 \gamma.
\end{equation}
This result is  natural: the larger    the increment of instability of linear waves $\gamma$, the larger norm $A_{0,M}$ is necessary for nonlinear modes to    overcome the instability of   linear waves and get born; in a similar way, the larger mismatch is necessary between   propagation constants of nonlinear and linear modes.}

In Fig.~\ref{fig:schematics} we also illustrate the behavior of the normalized asymmetry parameter $\cZ$ defined as
\begin{equation}
\cZ = \frac{Z}{A_0} = \frac{|a_R|^2 - |a_L|^2}{|a_R|^2 + |a_L|^2}.
\end{equation}
Simple  calculation gives the dependence of the normalized asymmetry measure $\cZ$ on  $A_0$:
\begin{equation}
\cZ^2 = \frac{1}{2}\left(1 \mp \sqrt{1 - \frac{A_{0,M}^2}{A_0^2}}\,\right),
\end{equation}
where $A_{0,M}$ is the minimal value of $A_0$ which defined in (\ref{eq:M}). Here   the minus sign corresponds to $A_0$ decreasing  from $+\infty$ down to its minimal possible value $A_{0,M}$, and, respectively, $\cZ$ increasing from 0 up to $1/\sqrt{2}$. The plus sign  corresponds to $A_0$ increasing from   $A_{0,M} $ up to $+\infty$, and, respectively, $\cZ$ further   increasing from  $1/\sqrt{2}$ up to its  asymptotic maximal value equal to unity.  In Fig.~\ref{fig:schematics} we use arrows to indicate the directions along the plotted curves that correspond to the increase of asymmetry parameter $\cZ$. Since   $\cZ$ is different from zero for all  solutions, we conclude that  either upper and lower subbranches  consist  of asymmetric  modes, that is $|\psi(x)|\ne |\psi(-x)|$ for all stationary envelopes $\psi$. \textcolor{black}{At the same time, the  modes that correspond to   lower subbranches   are   more asymmetric, because the corresponding asymmetry parameter $\cZ$ is larger as compared to   upper subbranches.}
 
Apart from the fixed-point stationary solutions,   two-mode system (\ref{eq:aRaL})   predicts a variety of stable closed orbits which, in terms of   full equation (\ref{eq:nls}), correspond to localized modes whose intensity $|\Psi(x,z)|$ changes periodically along the propagation distance.

\begin{figure}
	\begin{center}		
		\includegraphics[width=0.999\columnwidth]{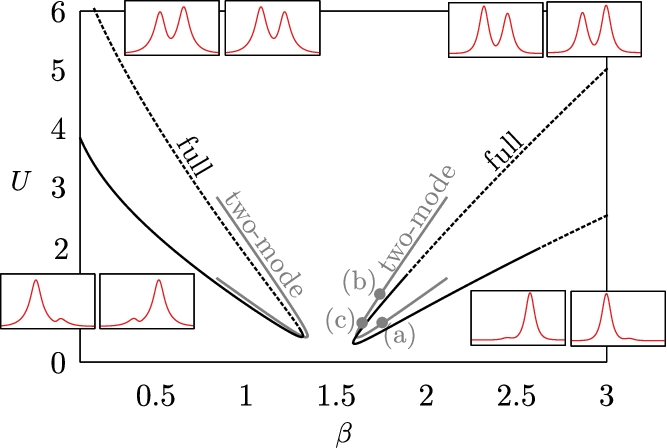}	
		\caption{Families of stationary nonlinear modes of full Eq.~(\ref{eq:nls}) (black  curves) on the plane $(U, \beta)$, where $U$ is the energy flow and $\beta$ is the propagation constant, juxtaposed with the analogous dependencies obtained from the fixed points of the two-mode system (gray curves). Solid and dashed segments of curves for the full equation correspond to stable and unstable stationary modes, respectively.  Small panels next to the energy flow curves illustrate \textcolor{black}{schematically} spatial shapes of nonlinear modes $|\psi|$ at the corresponding subbranches: weakly and strongly asymmetric states are situated at the upper and lower subbranches, respectively.   Points (a,b,c) at the two-mode curves in the focusing medium correspond to solutions used for dynamical simulations in Fig.~\ref{fig:focusing}. This figure is obtained for potential given by (\ref{eq:exp}) with $W_0=2$ and $\ell=2$. Nonlinear modes obtained  for focusing  ($\sigma =1$) and defocusing ($\sigma=-1$) nonlinearities are displayed simultaneously. \label{fig:families} }
	\end{center}
\end{figure}

\subsection{Stability restoration in the full equation}

To check the predictions obtained from the   simple two-mode system, we have computed  stationary states of    full equation (\ref{eq:nls}).  These have been sought in the form $\Psi(x,z) = e^{i\beta z} \psi(x)$, where $\beta$ is   real  propagation constant   and $\psi(x)$ is   stationary envelope. Numerical search of stationary states has been performed using a modified shooting approach adapted  for   peculiar properties of Wadati potentials \cite{KonZez14,Zezyulin2023}. The main results are presented in Fig.~\ref{fig:families} as dependencies of the energy flow (or squared norm) $U=\int_{-\infty}^\infty |\psi(x)|^2 dx$ on the propagation constant $\beta$. In the same figure we plot the analogous dependencies computed from     fixed points of the two-mode system. We observe a reasonably good qualitative agreement  between the results produced from the full equation and  from the two-mode  reduction. As the power flow becomes large enough, families of stationary modes get born  in either focusing and defocusing media through fold-like bifurcations. It should be stressed that  localized modes with real propagation constants form continuous families on the plane $(U, \beta)$ --- this is a special    feature of   Wadati \cite{KonZez14} and $\PT$-symmetric  \cite{Musslimani} potentials  as compared to  complex-valued potentials of other shapes.

\textcolor{black}{As the power flow $U$ increases, the quantitative agreement between the analytical and numerical   $U(\beta)$-curves  becomes worse. This is not surprising, because the two-mode analysis uses the linear eigenfunctions  as a basis to represent the nonlinear solution. Due to the growth of discrepancy,  we display only  lower  parts     of the analytical curves for nonlinear 	families in Fig.~\ref{fig:families}.}

All fixed points of the   bimodal  system are stable. However  this does not yet guarantee that   corresponding nonlinear modes of the full equation are stable too. We have therefore performed  a numerical stability check for stationary solutions of Eq.~(\ref{eq:nls}). It  proceeds in a standard way by considering a perturbed envelope $\Psi(x,z) = e^{i\beta z}[\psi(x) + u(x)e^{i\lambda z} + v^*(x)e^{-i\lambda^* z}]$, linearizing Eq.~(\ref{eq:nls}) with respect to small perturbations $u$,  $v$, and computing the instability increments given by imaginary  parts of eigenvalues $\lambda$. Fragments of energy flow curves that contain   stable and unstable solutions are shown, respectively, with solid and dotted lines in Fig.~\ref{fig:families}. \textcolor{black}{ We  observe  that in the focusing medium both   subbranches are stable close to the fold-like bifurcation, but become unstable for sufficiently large energy flow $U$. In the defocusing medium the lower (i.e., the strongly asymmetric) subbranch is totally stable while the upper one is unstable.  Therefore, even though the linear modes of the system are unstable, the stability restoration takes place for asymmetric nonlinear modes  in  both focusing and   defocusing media.}

\subsection{Dynamics of nonlinear modes}

\begin{figure}
	\begin{center}		
		\includegraphics[width=0.999\columnwidth]{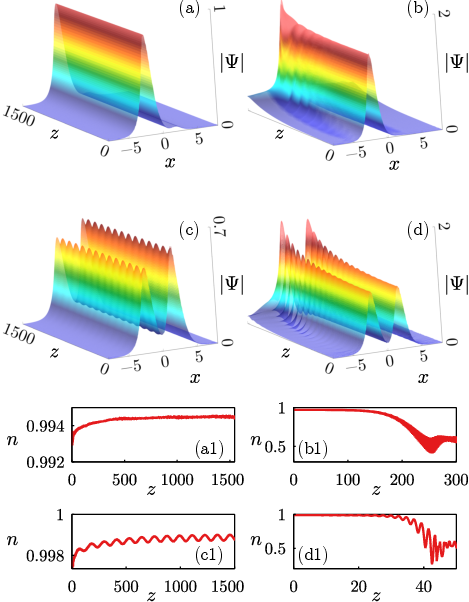}	
		\caption{Dynamical solutions of full Eq.~(\ref{eq:nls}) for   initial conditions obtained from   slightly perturbed  stationary modes $\psi(x)$ \textcolor{black}{with relative perturbation amplitude $\eta=0.01$}, under the focusing nonlinearity with $\sigma=1$: (a) lower subbranch, $\beta=2$ (stable); (b) lower subbranch, $\beta=3$ (unstable); (c) upper subbranch, $\beta=1.7$ (stable); (d) upper subbranch, $\beta=2.3$ (unstable). Four lower panels (a1)--(d1) display the corresponding dependencies $n(z)$, i.e.,  the energy flow fraction stored in both left and right modes as   defined by Eq.~(\ref{eq:n}).    \label{fig:stat_dyn_focus} }
	\end{center}
\end{figure}

Examples of nonlinear dynamics computed  from Eq.~(\ref{eq:nls}) with initial conditions taken in the form of stationary modes perturbed by a small-amplitude   noise are shown in Fig.~\ref{fig:stat_dyn_focus} (this figure corresponds to the focusing nonlinearity). \textcolor{black}{In presented simulations, the numerical  stationary envelope $\{ \psi_j\}$  computed  on the finite grid $\{x_j\}$, $j=1,\ldots, N$,   was perturbed by replacing $\psi_j$ with   $[1 + \eta(r_j  + i s_j) ]\psi_j$, where $0<\eta \ll 1$ is small relative  amplitude, and  sequences    $\{ r_j \}$ and $\{ s_j\}$ were generated as vectors of pseudorandom numbers  drawn from the standard normal distribution.}
In each case the dynamical behavior agrees with the linear stability prediction. In the same Fig.~\ref{fig:stat_dyn_focus} we additionally show  fraction of the energy stored in the left and right eigenfunctions  which is computed  as  
\begin{equation}
\label{eq:n}
n(z) = (|\tilde{a}_L(z)|^2 + |\tilde{a}_R(z)|^2)/U(z),
\end{equation}
where $U(z) = \int_{-\infty}^\infty |\Psi(x,z)|^2dz$ is the total energy flow, and coefficients $\tilde{a}_{L,R}$ are obtained by projecting the  soliton  onto the right and  left    states:
\begin{equation}
\label{eq:taRL}
\ta_{R,L}(z) = \int_{-\infty}^\infty \Psi^*(x,z) \varphi_{R,L}(x) dx.
\end{equation}
By definition, $0\leq n(z) \leq 1$, and   $n =1$ corresponds to the situation when the total   energy flow is perfectly stored in the two modes. For stable evolutions in Fig.~\ref{fig:stat_dyn_focus}, we observe that $n(z)$ remains close to unity, while unstable dynamics is   accompanied by excitation of modes that are  not taken into account   by the reduced  model.

\begin{figure}
	\begin{center}		
		\includegraphics[width=0.999\columnwidth]{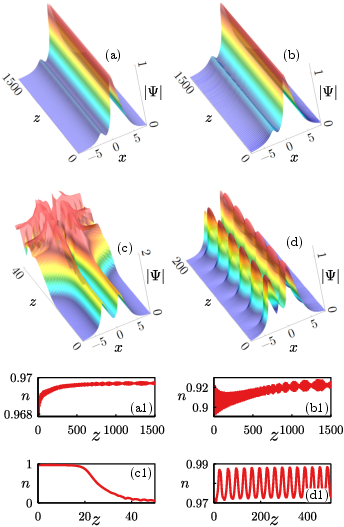}	
		\caption{Dynamical solutions of full Eq.~(\ref{eq:nls}) for   initial conditions obtained from   slightly perturbed  stationary modes $\psi(x)$ \textcolor{black}{with relative  perturbation amplitude $\eta=0.01$}, under the defocusing nonlinearity  with $\sigma=-1$: (a,b) lower subbranch, $\beta=0.6$ and  $\beta=0.2$ (both stable); (c,d) upper subbranch $\beta=0.6$ for \textcolor{black}{perturbation with   $\eta = 0.01$ (c) and $\eta=0.025$   and 5$\%$ (d)}. Four lower panels (a1)--(d1) display the corresponding dependencies of the energy flow fraction  stored in left and right modes as   defined by Eq.~(\ref{eq:n}).    \label{fig:stat_dyn_defocus} }
	\end{center}
\end{figure}

Examples of stable and unstable dynamics for stationary modes under the defocusing nonlinearity are shown in Fig.~\ref{fig:stat_dyn_defocus}. In accordance with the linear stability predictions, solutions from the lower subbranch  are stable, while those from the upper subbranch  are unstable. We have found  that behavior of unstable modes can be sensitive to the initial perturbation: for different   amplitudes of   random noise    perturbed  initial conditions can feature    extreme broadening [Fig.~\ref{fig:stat_dyn_defocus}(c)] or transform to an oscillating state [Fig.~\ref{fig:stat_dyn_defocus}(d)].  In the latter case the fraction of energy stored in the left and right modes remains close to   unity [Fig.~\ref{fig:stat_dyn_defocus}(d1)]. This indicates that a perturbed  unstable stationary mode   dynamically transforms into a stable solution corresponding to a closed orbit in the reduced bimodal system.

\begin{figure*}
	\begin{center}		
		\includegraphics[width=0.7\textwidth]{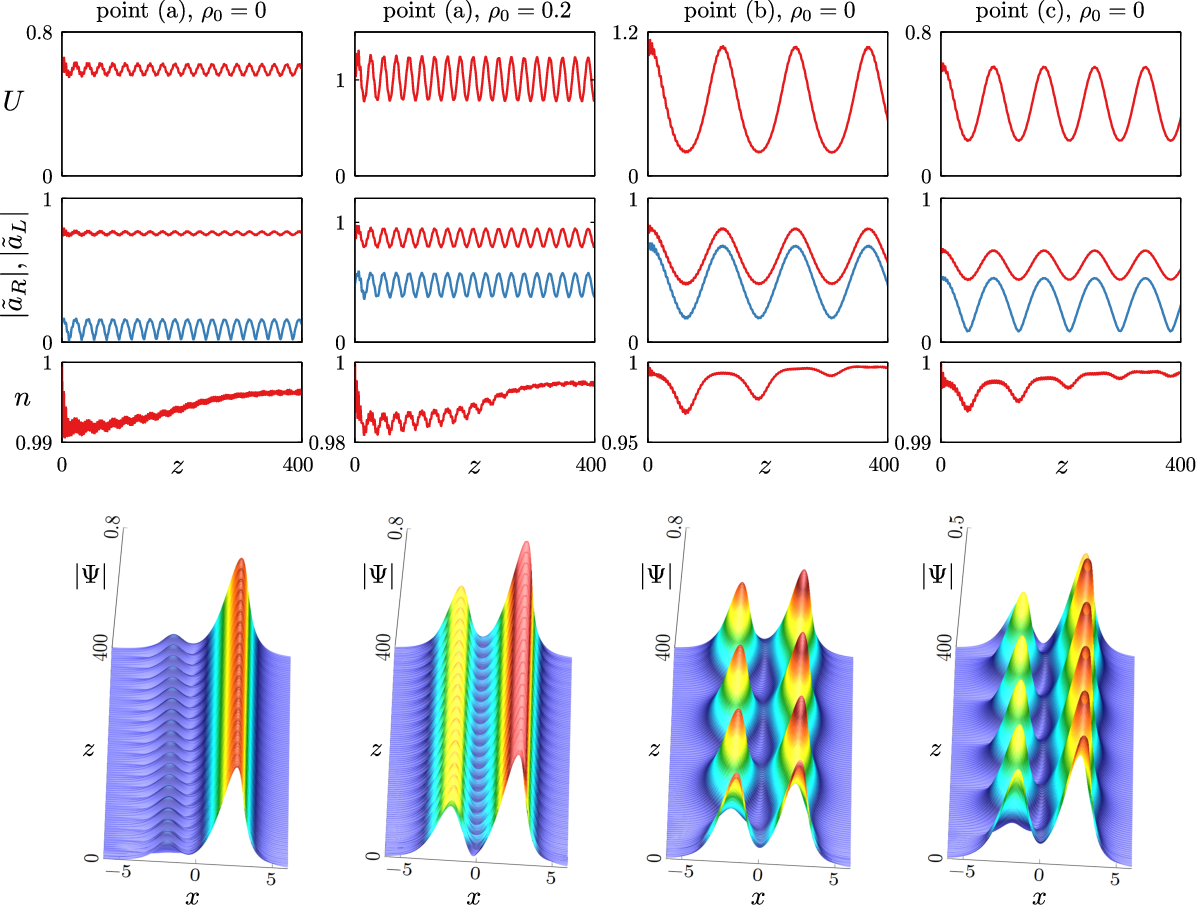}%
		\caption{Examples of evolutions computed from  full Eq.~(\ref{eq:nls}) with initial data obtained as fixed points ($\rho_0=0$) and periodic orbits ($\rho_0\ne 0$) of the  two-mode reduction.  Panels in the first, second, and third row show the  energy   flow $U(z)=\int_{-\infty}^{\infty} |\Psi(x,z)|^2dx$, amplitudes $|\ta_{R,L}(z)|$ of projections of the solution  $\Psi(x,z)$ onto the left and right eigenmodes [see Eq.~(\ref{eq:taRL})], and the fraction of the energy   conserved in the two modes $n(z)$ defined in Eq.~(\ref{eq:n}). Points (a,b,c) referred to in the upper row correspond to those labeled in Fig.~\ref{fig:families}. Full dynamical plots are shown in the bottom row.  (b) This figure corresponds to the focusing nonlinearity.  \label{fig:focusing} }
	\end{center}
\end{figure*}

\subsection{Periodic dynamics from the two-mode approximation}

The two-mode approximation developed above can also be used for   systematic   generation of  initial conditions that feature   nontrivial dynamics in the full equation. We illustrate this result using three points labeled as (a,b,c) in Fig.~\ref{fig:families}: point (a) belongs to the lower subbranch of the two-mode curve, and points (b,c) belong to the upper subbranch. We use the corresponding fixed points and periodic orbits of the reduced two-mode system to prepare   initial conditions for   full equation (\ref{eq:nls}). The results are shown in   Fig.~\ref{fig:focusing} as  dependencies of the energy flow $U(z)$ (upper row), amplitudes  $|\ta_{R,L}(z)|$ of projections of the solution onto the right  and  left  states [see the definition in (\ref{eq:taRL})], and the fraction of   energy    stored in the two modes  defined in (\ref{eq:n}).

We start with   point (a) and use the initial conditions corresponding to a fixed point. In terms of the two-mode solution this corresponds to the zero integration constant: $\rho_0=0$.   The energy flow and coefficients $\ta_{R,L}$ computed from the full numerical solution (and shown in the first column of Fig.~\ref{fig:focusing}) feature  oscillations of relatively small amplitude and therefore agree well enough with the predictions of the two-mode system (since the considered initial conditions correspond to the  fixed point, the two-mode system obviously predicts that the plotted dependencies must be  constant). Further, we increase   the integration constant by taking   $\rho_0=0.2$. In terms of the two-mode system this choice corresponds to a periodic orbit. The numerical solution obtained from the corresponding initial data (and shown in the second column of Fig.~\ref{fig:focusing}) indeed features more appreciable oscillations than the previous solution. However for points (b,c) presented in third and fourth columns of Fig.~\ref{fig:focusing}  we observe that     initial conditions corresponding to fixed points with $\rho_0=0$ develop in full solutions with   strong oscillations: in other words,  instead of   stationary modes corresponding to   fixed points we excite   solutions whose amplitudes $|\Psi(x,z)|$ periodically change  along the propagation distance. Even though these dynamics disagree with the predictions   of the bimodal reduction, the latter still remains useful, for   it enables    systematic preparation of initial conditions that develop in stable  periodic   solutions of full equation (\ref{eq:nls}).  
Oscillations of $\ta_{R,L}(z)$ obtained from     numerical solutions $\Psi(x,z)$ are always   inphase, which agrees with the prediction of the reduced system.  We additionally note that for all   data presented in Fig.~\ref{fig:focusing} the energy flow is almost perfectly stored in   left and right modes with $n(t)\geq 0.96$.

\section{Conclusion}
 \label{sec:concl}
We have demonstrated  a phenomenon of nonlinearity-induced stabilization for  a class of non-Hermitian optical potentials whose real part has   a  double-lobe structure. In contrast to   most of  previous studies where a similar behavior has been encountered, our system is parity symmetric  but not parity-time symmetric. Another salient  difference from earlier results is that the stabilization occurs for asymmetric nonlinear states that do not respect the parity symmetry inherent to the potential. Moreover, the stabilization takes place for either sign of   cubic nonlinearity. Analysis  of the stability restoration  has been  developed using  a simple two-mode system  obtained by projecting the solution onto superpositions of  linear eigenmodes    centered in   right and left wells of the potential. Dynamical simulations of full envelope equation confirm the existence of stable  nonlinear modes and  reveal the ubiquity of oscillating patterns with intensity  periodically changing  along the propagation distance. In contrast to   familiar  oscillations in   double-well potentials that  are most usually accompanied by the tunneling between the wells,  periodic patterns that we observe feature   distinctively different behavior with  the energies stored in the left and in the right wells oscillating inphase.  

\textcolor{black}{
Our results have been obtained for a special class of complex potentials that correspond to the form $w^2(x) + i w_x(x)$, where $w(x)$ is a real-valued and antisymmetric function and $w_x(x)$ is its derivative. In a real-world system the optical landscape will never perfectly match this shape, and the propagation of light will be governed by a perturbed      potential $w^2(x) + i w_x(x) +  \varepsilon r(x)$, where $r(x)$ describes    a complex-valued perturbation and  $0<\varepsilon  \ll 1$. Let us first address the situation when the perturbed  potential is still  Wadati-type, i.e., it can be represented as $y^2(x) + iy_x(x)$ for some $y(x)$ which is real-valued, but not necessarily antisymmetric. In this case   the effect of the perturbation is expected to be merely quantitative, and families of stable nonlinear modes are expected to exist. The difference between $y(x)$ and $w(x)$ will only result in  a slight deformation of     solutions.  However, if the perturbed potential does not belong to the Wadati class, then the effect of the perturbation becomes   stronger. For this case we expect that  stationary nonlinear modes predicted in our study will transform into `pseudomodes', i.e.,    approximate solutions with propagation constants  having small imaginary parts (of order of $\varepsilon$). In the   mathematical sense, those   pseudomodes are  not  authentic solutions of the governing envelope  equation. However,   these objects are expected to be dynamically robust, i.e., to  feature  nearly stationary evolution for sufficiently long propagation distances. Therefore they  can still be regarded as meaningful physical entities, and their quasistable behavior is expected to be  distinctively different from the authentically unstable propagation of linear waves. Some properties of such pseudomodes in a different class of complex potentials have been recently discussed in Ref.~\cite{ZSA}. }

\section*{Acknowledgments}
The research was supported by the Priority 2030 Federal Academic Leadership Program.

\end{document}